# VORTEX FLOW IN THE TECHNOLOGY OF RADIATION WAVE CRACKING (RWC)

L.A. Tsoy *, V.N. Kolushov, A.G. Komarov, A.N. Tsoy

(Heavy Oil Tech LLC)

RWC technology performs the process of physical and chemical modification of the heterogeneous organic crude stock into a useful demanded product with high efficiency and virtually in one step. For the first time the technology has been presented at the $2^{nd}$ Eurasian conference on nuclear physics and its application in Almaty in 2002 [1]. Gas used in this method is utilized into the liquid product.

RWC technology is as follows: the original oil is sprayed into the gas stream is fed into the reactor in the form of mist, where it is treated jointly by the accelerated electrons and the microwave radiation. Next, the irradiated crude stock is sent to a separator for separating the gas and claimed and unclaimed products. The unclaimed liquid and gas return to the chamber for repeated treatment by radiation with the flow of fresh oil and gas, and the claimed high quality products in the form of gasoline, kerosene and other products are withdrawn from the installation.

The experimental RWC installation was assembled on the basis of the reactor, combined with the linear accelerator ELV (up to 2.5 MeV, 150 kW) and a standard UHF generator (0.5 - 2.45 GHz., Up to 2 kW). A beam of accelerated electrons has a diameter of ~ 20 mm. It goes through the titanium foil into the reactor chamber with the heterogeneous medium.

Gas-liquid medium has a density ranging from about 1 to 0.1 g/cm3 (bubbles in a liquid, jet in the gas) and from 0.1 to 0.01 g/cm3 (drops in the gas). Hydroge, light alkanes, natural or associated gas can be used as the gas phase. The liquid phase is the heavy hydrocarbons, products of petroleum refining or the solutions of the chemical compounds (for example, carbon powder).

## 1. The experimental RVC installation.

Vortex flows are widely used in gas-liquid installations for generation of cold or heat, in the wind tunnels in aviation, astronautics and plasma processes.

The theory of vortex flows in fact explains in the new way the processes in the astronomical objects, in the atmosphere and in the microcosm. Therefore, it is understandable that hundreds of scientific papers are published during the year on this topic in the world, however the chemical processes in vortex flows are studied little. Many methods of chemical production, not excluding the widely used traditional thermocatalytic cracking owe their success to the use of such vortex flows.

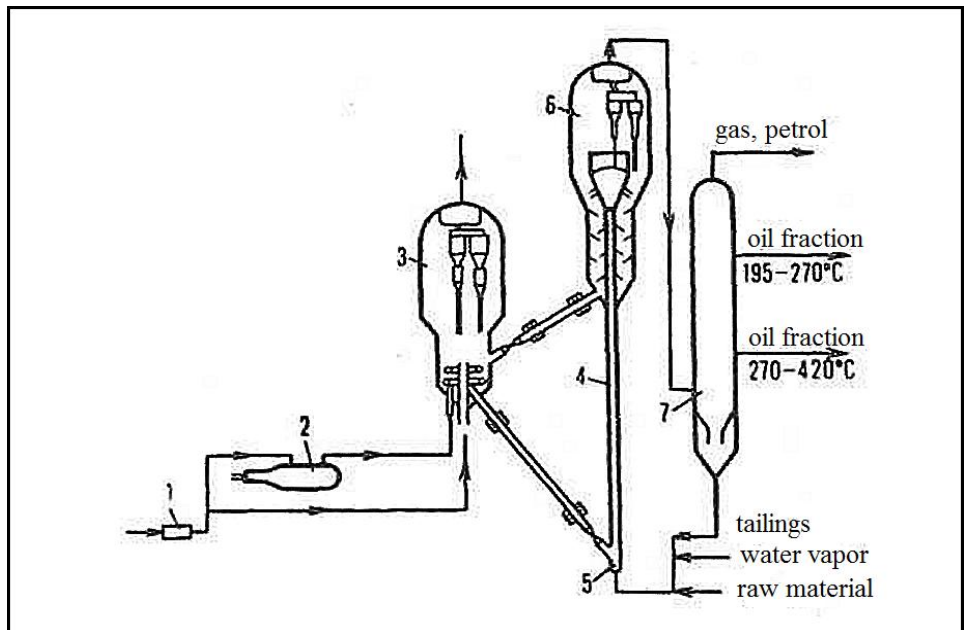

Fig.1. Cracking installation: 1-compressor, 2-vessel, 3-regenerator, 4- lift-reactor, 5-spray unit, 6-desorber, 7-distillation column

Fig. 1 shows an ordinary circuit of the most common thermocatalytic cracker installation with the micro spherical catalyst. Warmed in a tube furnace crude stock is fed into the so-called capture unit, where it is finely atomized and meets with a stream of regenerated catalyst. The process of splitting of the molecules is carried out in the lift-reactor. The reactor is in a form of steel tube lined from the inside. Its length is 30 -35 m in order to ensure the contact of the crude stock with the catalyst. The catalyst moves upwards in form of the crude stock vapor at a linear flow velocity of 10 - 12 m/s. At such velocity the flow is turbulent and definitely of swirl-type. The cracking product steam and the remainders of the crude stock after separation from the catalyst in the cyclone separators located in the reactor are separated in a distillation column [2].

Please note that the science has not yet developed a physical theory of catalytic cracking, and the processing technology is the result of a century of practical findings. The pervasive nature of the cracking technology makes the developers to carefully analyze these developments and look for ways to overcome the shortcomings, the most important of which is again the scale as in the multitude of processing steps and the high fire danger.

In our RWC installation [3] instead of heating the crude oil used is sprayed either by nozzle or by the ultrasonic nebulizer. The study shows that except for reactor all other RWC components assembled using the ready to use industrial devices, components and equipment. Based on the results obtained from the experimental installation, the optimal parameters of the commercial industrial plants for processing and transportation of heavy oil are determined.

Industrial RWC plants can be of two types: small on a mobile basis with the capacity of up to 100,000 tons per year and the relatively small stationary plant with the capacity of up to 1.2 million tons of crude oil. The product of industrial plant in form of a motor fuel can be used by the consumer with little additional refinement. If the product can be transported, then the RWC products can be sent to the large refineries for deeper processing into the staple products. The RWC process occurs almost at normal temperature and pressure, which is a huge advantage over other methods of refining. The RWC process control and operation can be fully automated, which is advantageous from the radiation and fire safety point of view.

Schematic diagram of the RWC installation is shown in Fig.2. The main components are: reactor, liquid phase circulation system, gas circulation system, linear electron accelerator, UHF generator, separator for withdrawal of claimed products.

Parameters determining the process are the ELV power and UHF radiation, the radius of the suspended particles, the forward and return gas flow velocity. The product yield depends on the depth of the crude stock processing and the separation method of the claimed and unclaimed radiolysis products.

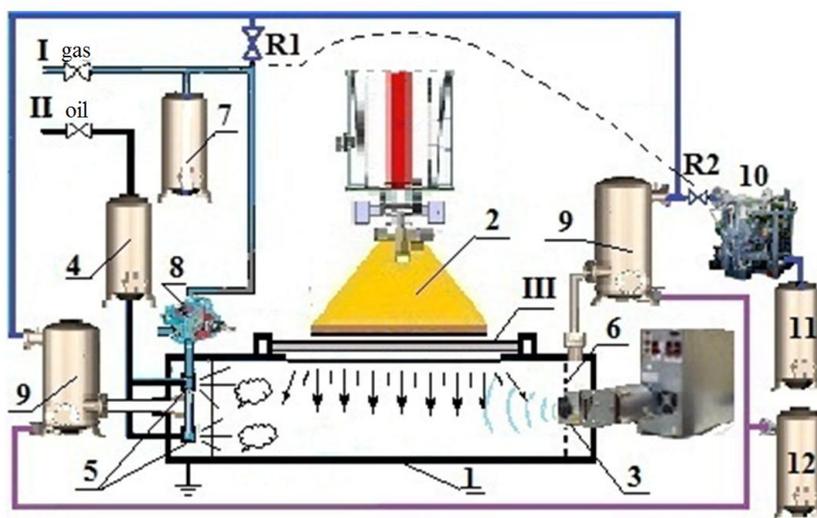

Fig.2. Schematic diagram of the RWC installation: 1-reactor; 2–linear accelerator ELV; 3- UHF generator; 4- crude oil vessel; 5 –nozzles; 6- flow rectifier; 7-receiver; 8- axial blower; 9- separators; 10- compressor; 11- gas cylinder; 12 – claimed products receiver; III-emergency valve; R1,R2 – simultaneous valves

The reactor is a steel pipe of diameter 400 - 600 mm and 1.8 - 2.4 m long, with the window for inlet of the accelerated electrons from the titanium foil and the waveguide of the UHF radiation. The liquid crude stock vessel is installed above (one floor up from the premises of the reactor), so the crude stock enters into the reactor by gravity. The gas supply system consists of a receiver and an axial compressor of the gas stream. Sprayed oil from the cavitation zone sweeps swirling through the ELV impact zone and UHF radiation, gets into the flow rectifier and goes to the separator through outlet sleeve. The outlet sleeve diameter is selected so that a central flow is formed in the opposite direction. The flow switches of the axial compressor and gas compression compressor work simultaneously in two modes. In a research mode, the oil aerosol is subjected to a one-time exposure, all products are disposed separately. In a continuous production mode only the claimed products are withdrawn from the system, and the rest returns to the reactor for repeated exposure together with the fresh portions of oil and gas. Note that the waste is almost zero, and the product volume increases compared with the initial volume on the account of the associated gas utilization (up to ½ of the weight).

Aerated oil after the nozzle is picked up by the gas flow directed along the tangent to the circumference of the inner surface of the reactor in order to twist it into a spiral. Passing by the accelerator window, the crude stock is exposed to the electrons, and approaches another end of the pipe, where the generator waveguide emits the ionizing radiation of the UHF field. Thus, the reactor can be divided into three zones of treatment. In the first zone, the crude mixture is transformed into the claimed product due to the cavitation bubbles and droplets, and the cavities occur in the oil layer flowing down the inner wall of the reactor. In the second zone of electron treatment the crude stock is exposed to the irradiation by the accelerated electrons to form ions, excited molecules, radicals and molecular compounds. The electrons undergo a significant change of direction, accompanied by electromagnetic radiation. In the third zone of UHF treatment the axial cold flow is formed, accompanied by the change of the kinetic energy and rotational moments of the flow. The division into power treatment zones is nominal, because the birth and collapse of bubbles occurs throughout the whole volume, the individual electrons "run" up to 1.5 m from the center the window, UHF radiation effects the entire volume of the reactor.

The RWC plant was used for study of the function of formation of the radiolysis products from the irradiation conditions, when the heavy hydrocarbon molecules are destructed and the gas molecules are enlarged into light fractions of motor fuel. This allows us saying that the radiation-chemical processes can be managed with 100% conversion of crude stock into claimed products.

Under the UHF influence and ionizing radiation the feedstock and gas molecules are destroyed with the formation of products containing light fractions of gasoline and kerosene with low molecular weight. This result follows from the analysis of spectra, chromatograms of the original Akshabulak crude oil and the product obtained in the result of irradiation in the RWC reactor (Fig. 3). The table shows real change in the fractional composition due to irradiation. It is clear that RWC results with deep factional change towards light motor fuel. Comparison of the infrared spectra before and after the radiation wave-cracking (Fig. 4) shows that irradiation by the ionizing radiation and electromagnetic field the products accumulate certain chemicals (e.g. aromatic hydrocarbons, with the wave number of 674.82-benzene, 693.83 toluene, 727.01, xylene, etc.). Certain direction of chemical transformation allows obtaining the products with desired properties subject to the irradiation conditions. Shifts of the fractional composition change the physical and chemical parameters of crude oil. The viscosity of the product amounted to 1.3 mm$^2$/s, i.e. decreased by 10 times, and pour point decreased from +21° C to -21°C.

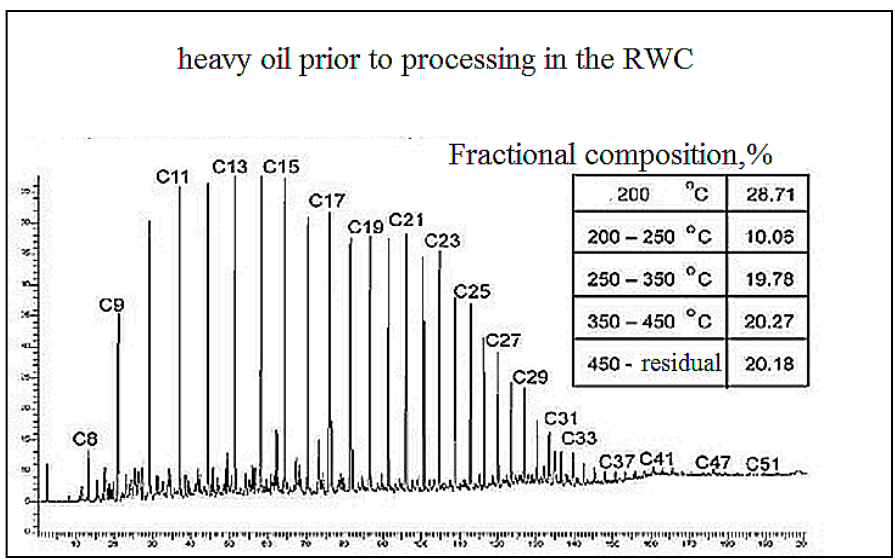
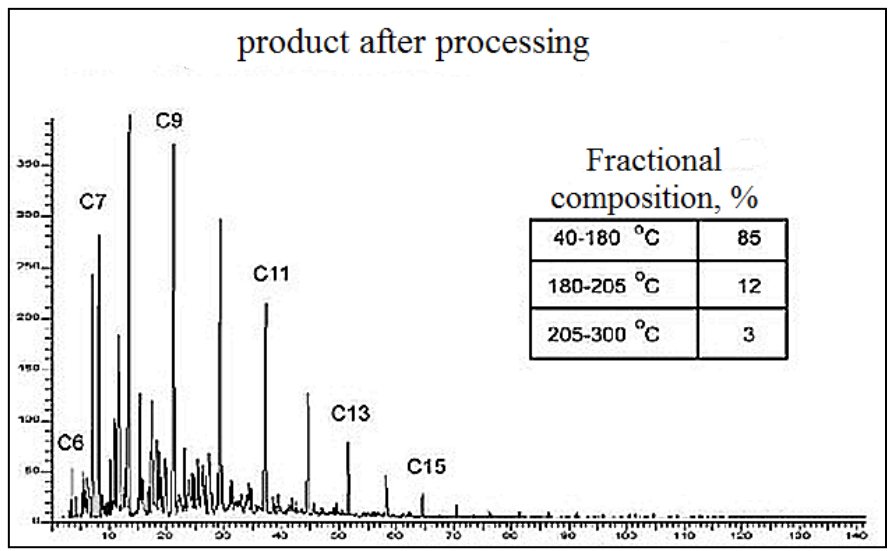

Fig. 3,4. Chromatogram of heavy oil and the radiolysis product from the RWC installation

The resulting unstructured oil was injected into the transported oil in the amount equal to 20% on the amount of transported oil. The viscosity of the transported crude oil decreased by 2 times and the pour point went down to 0°C on the account of dissolving the high-molecular substances contained in the transported heavy oil.

## 2. RWC processes modelling.

There is still no clear scientific explanation of why the reverse axial flow appears in the vortex flow when the temperature of the axial gas flow in the Ranke pipe (Fig. 5) falls down to - 200°C. The gas flowing at the velocity v has the thermodynamic temperature T* of the thermal motion of gas molecules and the cooling temperature T measured by a fixed thermometer placed in the flow path.

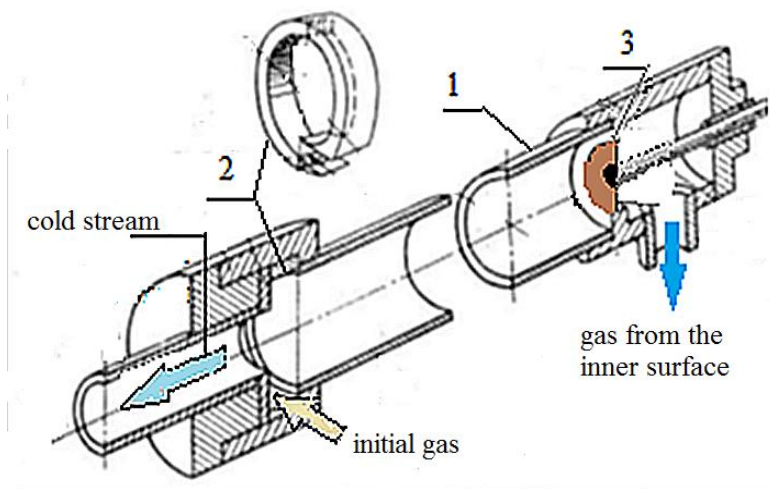

Fig. 5. Ranke vortex tube: 1- body; 2- Snail with the source gas, 3 - annular space to exit of the wall flow.

These temperatures relate as follows:

$T = T^* + v^2/2S_r$        (1)

where $C_p$ is the gas specific heat capacity.

From this relation it follows that with the acceleration of the flow to the velocity v in the adiabatic conditions, the thermodynamic temperature T decreases. And vice versa, with the decrease of the thermodynamic temperature under adiabatic conditions, the flow rate increases. The kinetic energy of the flow increases by:

$\Delta E = m \Delta v^2 /2 = m C_p (T - T^*)$     (2)

where m is the mass of the gas passing through the reactor.

Note that the vortex tube can work with any gaseous working solid and at the variety of differences of atmospheric pressure starting from fractions up to hundreds of atmospheres. The gas flow rate can vary from fractions of m$^3$/hr up to hundreds of thousands m$^3$/hr. With the increase of the pipe diameter the efficiency of the vortex flow increases. In general, the mathematical model must explain the whole process of radiolysis, as shown in the diagram.

Attempts to create a theory of vortex flows by constructing and solving the system of gas dynamic equations lead to insurmountable mathematical difficulties. More and more new peculiarities reveal, which do not fit into the usual physical concepts. V.E. Finko [4] obtained cooling down to 30°K in his experiments in an expanding pipe with the cone angle of 14°.

The effectiveness of the installation in the experiments increased with the increasing gas pressure at the inlet up to 4 MPa and above, which contradicts to the conventional theory. The temperatures of the hot and cold streams were significantly lower than the temperature T of the gas fed into its vortex tube. This means that the energy balance in it does not correspond to the well-known Hilysh balance equation:

$$I Q = I_x Q_x + I_r Q_r \qquad (3)$$

where I, Q - flow rate and enthalpy of the inlet gas

$I_x$, $Q_x$, and $I_r$, $Q_r$ - enthalpy of the used hot and cold streams of gas.

Hence, the cold stream portion must obey the relation:

$$Q_x = Q (T_r - T) / (T_r - T_x) \qquad (4)$$

where T, $T_r$ and $T_x$ – temperatures of inlet, hot and cold streams.

Experiments proved that the ratios (3 and 4) are not obeyed. The discrepancy between theory and experiment for the enthalpy of incoming and outgoing flows amounted to 9-24% and increased with the increasing inlet pressure or with the decreasing temperature of the inlet gas.

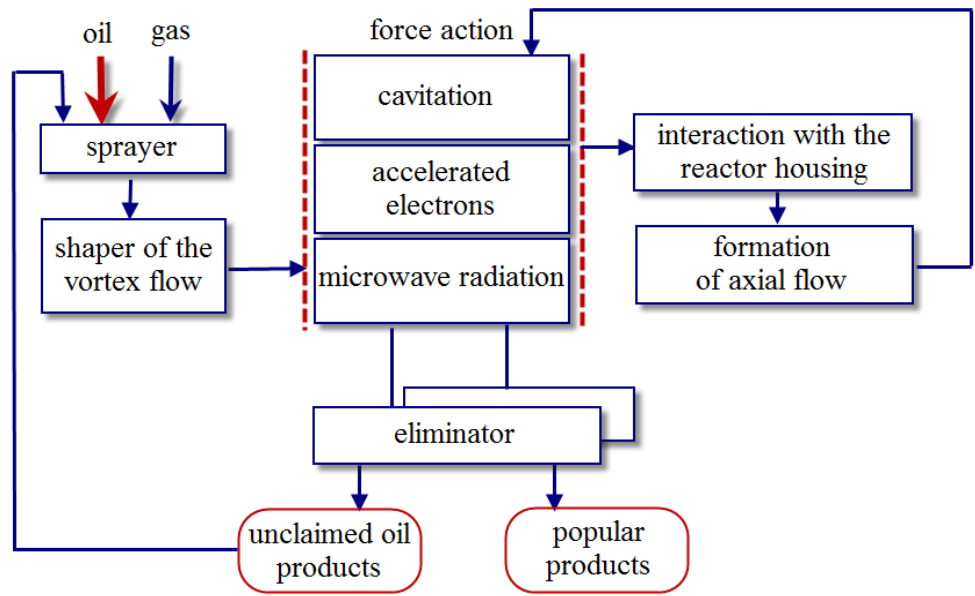

**Fig 5. RWC process diagram**

None of the previously proposed mechanisms of heat transfer, including the mechanism of turbulent heat exchange counter, does not explain the high rate of heat removal from the gas with the temperature drop of ~70°K or more.

The rod introduced along the axis of the vortex flow, another end of which is fixed to the bearing, rotates at 3000 rev/min in the direction opposite to rotation of the input vortex flow. This contradicted the established concept of the mechanism of formation of the axial flow, which according to all common views must rotate in the same direction as the main flow (the "hard disk". model). Another interesting fact was discovered: at the output of the cold gas stream an infrared emission of the band spectrum in the wavelength range of 5-12 microns appeared. Its intensity increased with the increasing gas pressure at the inlet pipe. Sometimes one could also observe a "blue light coming out of the flow core". Unfortunately, the researcher did not attach special significance to the light, noting it as a curious side effect. Y.S. Potapov noted this fact as important [5], and suggested that the cooling of the axial flow occurs due to radiation emission by the flow, which carries the energy away. The pipe walls absorb radiation and get heated up for this reason. The peripheral gas flow contacts with the pipe walls, withdraws the heat and warms up. With this mechanism, the reactor vessel serves as an intermediate body, providing heat transfer from the central to the peripheral flow. The quantity of radiation energy, calculated by the author using the virial theory, turns out to be large enough, exceeding twice the kinetic energy of the flow: $W = 2W_k$.

A number of researchers have shown that the large heat gradient on the border of the axial and annular flows stabilize the helical toroidal vortex. Lowering the temperature in the boundary layer can be calculated by the formula:

$$\Delta T = v^2/2C_p \qquad (5)$$

where v is the velocity of the boundary surface; Pr-Prandtl number, $C_p$ - specific heat capacity of the environment at the constant pressure. Prandtl number (a measure of the transfer of the molecular momentum to heat):

$$P_r = \mu \, C_p / \lambda \qquad (6).$$

where $\mu$ - dynamic viscosity; $\lambda$ - thermal conductivity of the medium.

It is estimated that the maximum velocity gradient at the surface of the vortex flow reduces the viscosity of the boundary layer to a minimum value. As viscosity decreases the energy transfer to the neighboring layers decreases, which leads to increased stability of the vortex formation.

One more structural detail should be noted: if one sets the "flow rectifier" at the "hot" end of the vortex tube, the temperature of the axial flow is not so low as without "rectifier". This gives rise to noise, and the "rectifier" gets very hot, which cannot be explained only by the friction of the gas stream with the "rectifier".

Inside the RWC reactor the multi-phase mixture moves, which is formed by spraying of oil into the stream. Hydrogen, gaseous alkanes (methane, butane, propane, etc.) and their mixtures, natural or associated gas can be used as the gas. In general, the flow is the fluid with inclusion of gas and solids with density of 1 - 0.1 g/cm$^3$, and the gas flow with liquid droplets and heavy hydrocarbon molecules of high-density 0.1 - 0.01 g/cm3.

The distribution of these substances inside the reactor has a complex configuration. Drops of liquid contact with the walls of the reactor, forming a liquid film flowing down, which interacts with the counter flow of gas. The presence of the film must be considered in the study of hydrodynamic flow of the mixture and its interaction with the counter-flow of the vortex flow.

When the flow gets steady a model of two-phase mixture can be used. The most important characteristics of this flow are the mass and volume fractions of phases, respectively, in mass and volume flow of the mixture. The volume gas concentration (volume gas content) will be determined by the expression $\varphi = S_1/S$, where $S_1$ - cross-section area occupied by the gas, S- cross-section area. The average concentration of the liquid phase is a fraction of the volume of the mixture occupied by the liquid. It will be expressed by the value of $1 - \varphi$. Density $\rho$ of the mixture can be determined as follows:

$$\rho = \rho_r \varphi + \rho_z h (1 - \varphi) \tag{1}$$

where the subscripts "r" and "z" will denote the parameters related to the gaseous flow and liquid droplets, floating in the stream.

The volume gas content differs from the volume consumption gas content due to the relative motion of the phases. Flow, volume concentration (gas content) is determined as follows:

2. $\beta = Q_r/(Q_r + Q_z)$ (2)

where Q - phase volume consumption.

Consumption mass concentration is related to the consumption volume concentration by the phase density

$$X = \rho_r Q_r / (\rho_r Q_r + \rho_z Q_z) = G_r/(G_r + G_z) \tag{3}$$

where Q – phase mass consumption.

Mass concentration of gas and liquid is as follows: (3)

$$C = \rho_r \varphi / \rho; \quad C_z = \rho_z (1 - \varphi)/\rho \tag{4}.$$

Dynamic viscosity of the mixture is given by:

$$\mu = \mu_r X + \mu_z (1-X) \tag{5}$$

where $\mu$ - viscosity of the phases.

Provided that $\varphi = \beta$ the droplet sizes are comparable to the molecular dimensions, and the flow can be considered homogeneous, and the velocity of gas and liquid droplets in the flow to be same. Such flow model is called quasi-homogeneous and one can talk about a two-speed flow regime, referring to the velocity of the mixture and of the liquid film.

However, more often the droplets exceed greatly the size of the largest molecules. Then the medium is no longer homogeneous and the three-speed model should be used. The heavy oil molecules are dissolved in all three phases. In the gas stream they easily separate from the liquid and acquire rotational motion under the influence of the convection forces.

Under the gravity the liquid partially precipitates to the bottom of the chamber. When the volume concentration of the gas $\varphi <0,3$ the flow regime of the gas-liquid mixture usually becomes bubbly. At the gas phase velocity v> 5-10 m/s, at which the structure is unstable, the liquid film starts appearing on the wall of the channel and the whole space accepts mostly the foam structure. With the increasing volume gas concentration $\varphi> 0,6 - 0,8$ the flow regime turns into film or annular, in which the liquid phase forms a continuous film flowing along the wall in the direction of the gas phase. Because of the dynamic interaction of the gas flow and the liquid film formed on the surface of the latter the waves appear, from which the small drops and bubbles are blown off and under the influence of the centripetal force of the vortex are displaced in the direction of the axial flow. When the gas-liquid mixtures flow in the vertical channels, then almost axisymmetric concentration distribution and the phase velocities over the cross section gets established for all flow regimes. When the mixture flows in the horizontal and inclined channels, then we have to take into account the effects of gravity that violate the axial symmetry of the phase distribution over the cross section.

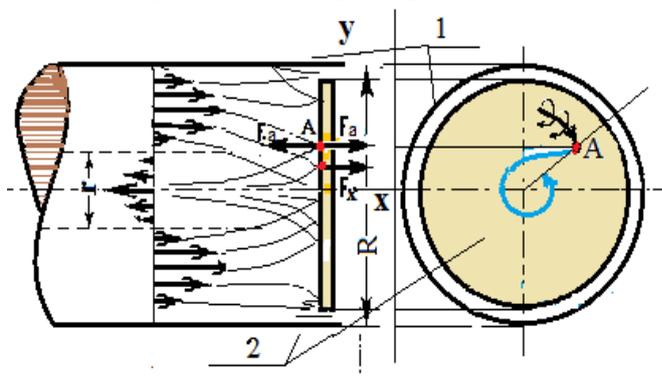

Fig.6. The profile of the velocities in the RWC reactor:
1-reactor vessel; 2- flow reflector

The physical cause of the axial flow in the Ranke pipe still has no explanation. A common view that it appears due to the centrifugal forces, which cast off the particles to the periphery, resulting in a dilution zone of the axis, and the air rushes to the front of the jet, cannot always be justified. Fig. 6 shows the velocity profile along the axis of the RWC reactor when there is an axial flow, which shows that the static pressure along the axis get the maximum at the inner pipe wall and on the border of the wall and axial flows, when the axial component of the velocity $V_x = 0$. Due to the rotation in the wall flow there is a centripetal force. Also there is a centrifugal force in the axial flow. These forces are directed opposite to each other, but contrary to Newton's third law are not equal. The centripetal force exceeds the centrifugal, which explains the stability of the vortex flow.

It is generally accepted that the reverse axial flow occurs when the output of the mixture through the end of the tube is partially blocked, and it creates excessive pressure. But the burner experts say that the backflow along the axis of a swirling jet occurs even in the absence of the walls of the apparatus.

In our experiments, the backflow together with the good quality gasoline appeared behind the fan in a narrow hose with the diameter of about 10 mm and more than 10 m long from the cylinder, which was located one floor above, where there could be no excessive pressure.

We dare to explain that the backflow appear in the swirling jets without referring to the theories of relativity, or movement in time. The flow of the gas mixture with the oil in this case passes through the power zone, and part of the flow passes freely through the outlet apperture. Significantly larger portion with the average velocity along the axis $V_x$ hits against the solid reflector. The law of conservation of momentum requires that if the vortex flow acquires an axial momentum, then some other body (for example, the vortex device body) simultaneously acquires the same absolute value of momentum in the opposite direction.

In the closed vortex devices, when there is no contact between the vortex flow and the walls of the device (as in some cases with the free swirling jets), then the axial portion of the flow is forced to acquire a reverse impulse, which has a lower tangential velocity than the peripheral flow. In case of absolutely elastic impact the force of the particles against the wall equals the counter-force $F_{-a} = F_a$, and the axial flow gains the momentum $P_{-a} = - (m - m_k) V_x^2$ (minus sign indicates a change of momentum in the opposite direction), and the kinetic energy $E = (m - m_k) V_x^2 / 2$ remains unchanged. Thus, the wall is impacted by the forces $F_x$ and $F_a$, which are equal and in the same direction. These are the forces, bringing the so-called "gratuitous energy" of the vortex flow which is transformed into heat of the bottom of the reactor by cooling of the axial flow. Hence, the value of additional energy in excess of the energy consumed for formation of the vortex, must satisfy the inequality:

$E_c \geq 2mvt (1 – S_k/S) \sin\alpha$
    (6)

where the $E_c$ is the energy absorbed by the bottom of the tube, t - time of interaction between a particle and a solid barrier, S – cross-section area of the reactor, $S_k$ – area of the outlet aperture at the end of the reactor, $\alpha$ - step of the vortex spiral.

The tangential velocity of the circular motion of the particles in the vortex $v_y$ determines the centripetal force $F_y = mv_y^2/R$, which is directed towards the axis of the system and determines the radius $r_0$ of the axial flow.

We used the inequality sign in the expression (6) for the following reason. None of the researchers of the vortex flow takes into account the fact that the particles in the vortex flow can rotate, that is, that they have an intrinsic angular momentum $L = I\omega$, where I-momentum of inertia, $\omega$ - angular velocity of rotation around its own axis, respectively, the energy of rotation. In an elastic collision of a rotating particle at point A against a solid wall the precession rule is valid according to which the opposition is directed perpendicular to the acting force, just in the direction of the axial direction. Example is a strike of a rotating tennis ball against the ground, when the angle of reflection is not equal to the angle of attack, and after hitting the ground the ball changes its direction of rotation to the opposite. Hence, when the particle hits the hull, the latter receives additional energy $2E_{vr}$ and the axial flow - direction of rotation opposite to the main stream. In view of (6) the total energy of the vortex flow is defined by

$$E = mv^2 + m_{ob}v_{ob}^2 + 2mvt(1 - S_k/S)\sin\alpha + L\omega_c^2 r_c^2 \quad (7)$$

where m is the total mixture mass, L is the moment of inertia of the rotating particle with radius $r_c$ and with an angular velocity $\omega_c$.

This relation is fully consistent with the theory of the virial, when the average energy in time of a bound system in absolute value by 2 times greater than the average in time total kinetic energy of the motion of these bodies relative to each other:

$$E_b = -2E_{kin} \quad (8)$$

The energy balance in a vortex tube (3) in general takes the form:

$$I Q = I_x Q_x + I_r Q_r + W_c + W_t \quad (9)$$

where $W_c$ is the energy of intrinsic rotation of the mixture particles hitting against the wall, $W_t$ - energy carried away to the media from the walls of the tube.

The relation is not final because it was obtained under the assumption of elastic collisions, and the probability of inelastic collisions has not been investigated.

In the experiments by V. Finko the expression (4) could not be valid just because of the fact that they did not take into account the rotational energy of the particle around its own axis, which reached a significant value of up to 20% of the total energy.

In case of interaction of the elementary particles and molecules one must consider also the spin fluctuations in the elastic and inelastic collisions. They may not be large, but they can change the speed of the chemical reactions by hundreds of times.

Along with the models (8) and (9), which will be refined, we have developed a model based on the regression analysis:

$$Q_K = \operatorname{minmax}\left(Q^K_o + \sum_{i=1}^{n} a^K_i X_i + \sum_{i,j=1}^{n} a^K_{ij} X_i X_j + \ldots + a^K_{1\ldots n} X_1 \ldots X_r\right) \quad (10),$$

where k = 1,2, ... - the expected number of fractions,

$Q_K$ – yield of the expected product of the k fraction,

$Q_{Ko}$ - the average level in the regression model k,

$a^k_i, a^k_{ij}, \ldots, a^k_{i\ldots n}$ - the regression coefficients of the model k,

$X_i$ - factors: $M_r$, $M_n$ - the mass of gas and hydrocarbons crude stock, kg;

$W_e$, $W_v$, $W_v$ – power of the absorbed dose of the accelerated electrons, electromagnetic and ultrasonic waves, J/kg;

v - velocity of the gas flow, m/s, etc.

As a result of the factor analysis by varying 11 process parameters we arrived at relation (10) with significant regression coefficients:

$$Q_{PR} = 3,8 + 11\ W_3 + 25\ W_v + 81\ W_e - 27, lv \qquad (11).$$

This is just an estimation model, because we did not target obtaining the accurate values of the regression coefficients. Therefore, instead of the absorbed doses we took the electric power consumed by the respective devices like ultrasound, the UHF generator and ELV-4. From these data we can conclude that the efficiency of the ultrasound impact, UHF and accelerated electrons at comparable power consumption are in the ratio 1:7:21. The least effective is cavitation, the UHF impact is quite significant, only 7 times less than the impact by the accelerated electrons.

Further research will determine the constants of the rate of the radiation-chemical reactions in the processing of heavy oil and the parameters of the energy balance in the RWC process.